\documentstyle[12pt]{article}

\def\title{
\bgroup%
\obeylines\everypar={\hskip\parfillskip}%
\large
\bf\vrule height1cm width 0pt\relax}
\def\endtitle{\vskip1sp\egroup}
\def\author#1{\hbox to\textwidth{\hss\vrule height.9cm width0pt\relax%
#1\hss}}
\def\contauthor#1{\hbox to\textwidth{\hss\vrule width0pt\relax%
#1\hss}}
\def\moreauthors#1{\hbox to\textwidth{\hss\vrule height.8cm width0pt\relax%
#1\hss}}
\def\instit{\bgroup\small\it\obeylines\everypar{\hskip\parfillskip}}
\def\endinstit{\vskip1sp\egroup}

\begin{document}

\vglue -4 true cm
\vskip -4 true cm
\begin{center}
{\hfill }{\Huge DRAFT}
\end{center}
\vskip 2 true cm

\hyphenation {conti-nuous}
\hyphenation {Re-normalization Renor-malization Renorma-lization
Renormaliza-tion}

\begin{title}
Renormalization Group Study of the soliton mass
on the $(\lambda \Phi^4)_{1+1}$ lattice model
\end{title}

\vskip3truecm

\author{J. C. Ciria, A. Taranc\'on}
\begin{instit}
Dpto. de F\'{\i}sica Te\'orica, Facultad de Ciencias,
Universidad de Zaragoza, C/ Pedro Cerbuna 12, 50009 Zaragoza, Spain
\end{instit}

\vskip2truecm

\begin{abstract}
   We compute, on the $(\lambda \Phi^4)_{1+1}$ model on the lattice, the
soliton mass by means of two very different numerical methods. First,
we make use of a ``creation operator'' formalism, measuring the decay of
a certain correlation function. On the other hand we measure the shift of the
vacuum energy between the symmetric and the antiperiodic systems. The obtained
results are fully compatible.

   We compute the continuum limit of the mass
from the perturbative Renormalization
Group equations.   Special attention is paid to ensure that we are working
on the scaling region,
where physical quantities remain unchanged along any Renormalization
Group Trajectory.   We compare the
continuum value of the soliton mass with its perturbative value up to one
loop calculation. Both quantities show a quite satisfactory agreement. The
first is slightly bigger than the perturbative one; this may be due to the
contributions of higher order corrections.

\end{abstract}

\newpage

\section {Introduction}

   Standard perturbation is known to be a useful tool for the formulation of
Quantum Field Theory starting from Classical Field Theory. It has, however,
serious handicaps such as the fact that non-perturbative effects are not taken
into account. An alternative possibility is the quantisation of non-trivial,
non-perturbative solutions to the classical equations, such as solitons.

  The study of such topologically non trivial vacua in Field Theories
pre\-sents
several problems when the model is formulated in the continuum or in the
lattice.
In the continuum it is very difficult to extract non perturbative quantities,
like mass, and in the lattice, where this is possible with Monte Carlo
simulations, other problems are present.

  First, on the lattice the analysis of this kind of configurations is made
difficult by the trivial topology of the lattice, and because the concept
of continuity is lost
\cite{Schierolz}.
Second, it is of fundamental importance to define
how to measure quantities on these topologically non trivial vacua
\cite{Evertz}. We can consider how to compute the mass, for instance.
As is known, solitons are characterized by a topological charge,
related to their behaviour when spatial coordinates tend to infinity
( $Q \sim \Phi(x \to \infty) - \Phi(x \to -\infty$) ). This charge is
conserved with time. When we quantise a soliton, we obtain a quantum
soliton-particle and a
series of excitations of this particle, a so called soliton sector. Topological
charge becomes then a quantum number characterizing the sector. Its
conservation
prevents the soliton from falling to the vacuum, ensuring its stability.
The standard way of calculating the mass is considering an operator
with non vanishing projection on
this sector, then computing the connected correlation to large distance,
and finally extracting the mass from the coefficient of the exponential decay.

In the general case, for topologically non trivial sectors, the definition of
such an operator is very ambiguous. It is possible to define many
operators on the lattice sharing the same continuum limit, although their
behaviour far from this limit differs from each other. On the lattice,
the region where we can obtain results within reasonable computation times
is generally far from the continuum limit -where very big sizes would be
necessary-, and all those continuum-equivalent operators give us different
results.
\cite{Monopolo}.

  In four dimensional theories computer limitations have made this point
particularly difficult.

Fortunately, some interesting facts can be studied
quite satisfactorily in less than four dimensions.
We consider on this paper the $(\lambda \phi^4)_{1+1}$ model, where solitons
are also present.

   On a finite lattice the boundary conditions fix up the topological sector.
Periodic conditions fix the trivial vacua, for instance. Antiperiodic
conditions fix  vacua with non trivial topology (if the symmetry is broken).
Only free boundary conditions allow us to have different topological sectors;
however the finiteness of the system allows us to travel between vacua, and we
finish always in the trivial sector, the energy of which is lower.

In this model it is possible to carry out the computation of the soliton mass
by using two different, related, methods.

   First, we have made use of the operator defined by Kadanoff et al.
 \cite{Kadanoff};
in Spin Systems, its effect can be seen as the introduction of a
twist: a topological excitation induced by a specific dislocation of the
lattice.   It has  a  topological charge different from zero. Consequently, we
expect a non-zero projection onto the soliton sector.

 On the other hand, we can consider
the system with antiperiodic spatial conditions for the scalar field. This
system can be considered as the periodic one after the introduction of
a twist along the whole lattice time. The difference of the energies
of the periodic and antiperiodic systems, which is a local, easy to
measure quantity, provides us with another method to compute the soliton mass.

   We always keep in mind that a theory in a lattice acquires physical
meaning  only when we make its spacing tend to zero. In order to get to the
continuum limit \cite{Wilson} we use the Renormalization Group (RG) equations,
which are known in this model. We must consider the limit of zero lattice
spacing. A change in this spacing, and a change in the coupling
constants in such a way that the physical observables remain unchanged,
can be carried out by using the RG equations.

   Iterating RG transformations, we obtain a series of points in the parameter
space, - Renormalization Group Trajectory (RGT)-, which can be
characterized by a parameter {\it l}.  Different points of a trajectory,
corresponding to different values of the parameters, are obtained after
integrating over successive
energy scales. Thus, when  we move on any RGT,
the Physics remains the same.  In this way,
for example, as we evolve on a RGT, we see different values of the
correlation length on lattice units,
$\xi_0(l)$, but this correlation in physical units
$\xi = a(l) \cdot\xi_0(l)$ remains constant.

\subsection{$\lambda \phi^4$ model}

   We study the $\lambda \phi^4$ model in  d=2 dimensions,
the euclidean lagrangian density of which is given by
\begin{equation}
{\cal L}_{\rm euc} = {1 \over 2} (\partial_{\mu} \Phi)^2 + { r \over 2} \Phi^2
+
{{\lambda} \over {4!}} \Phi^4 \quad,
\end{equation}
where $\Phi$ is dimensionless, [$\lambda$] = $l^{-2}$, [r]= $l^{-2}$.

  In order to adapt our lagrangian to the lattice,  we proceed as usual:
\begin{equation}
\int_0^L dx  \to   a^d \sum_{n=0}^{L \over a} \;;   \hskip2truecm
\partial_{\mu}\Phi \to  { {\Phi[(n+\mu) \cdot a] - \Phi[n\cdot a]}\over a},
\end{equation}

where $L$ is the lattice extension. We conclude
\begin{equation}
 S_{\rm euc} = - \sum_{n,\mu} \Phi_n \Phi_{n,\mu} + \sum_n
\{  (d+{{r_0} \over 2}) \Phi_n^2   +  {{\lambda_0} \over {4!}} \Phi_n^4\} \;,
\label{discrete}
\end{equation}
 We introduce the following notation: the adimensional parameters defined on
the lattice are subscripted; thus we use $\lambda_0,r_0 $ (respectively equal
to
$\lambda  a^2, r  a^2 $).

Making the spatial coordinates discrete implies imposing a momentum cut-off
$\Lambda = {{2  \pi} \over a}$. After scaling the momenta $q \to p =
{q \over \Lambda}$ we can express (\ref{discrete}) in momentum space
\cite{Wilson}:

\begin{equation}
S_{\rm euc} = {1 \over 2} \int_p (p^2 + r_0) \Phi(p) \Phi(-p) +
{{\lambda_0} \over {4!}}  \int_{p_1}\int_{p_2}\int_{p_3} \Phi(p_1) \Phi(p_2)
\Phi(p_3) \Phi(-p_1-p_2-p_3) \,,
\end{equation}
with $ \int_p \equiv
\int_0^1 {{d^d p} \over {{(2 \cdot \pi)}^d}}$.
  In d=2 we have two fixed points  \cite{Wilson}:

   i) The gaussian point, $S_{\rm gauss}= \int_p u_2^*(p) \Phi(p) \Phi(-p)$,
with
$u_2^*(p) \sim p^2$. That is to say, taking just the kinetic part
of the lagrangian.

   ii) A non-trivial point, which is built adding to the lagrangian the
term
\begin{equation}
S=\int_{q_1}\int_{q_2}\int_{q_3} u_4^*(q_1,q_2,q_3,-q_1-q_2-q_3) \Phi(q_1)
\Phi(q_2) \Phi(q_3)\Phi(-q_1-q_2-q_3) \;,
\end{equation}
with  $u_4^* (q_1...q_4)\sim (q_1^2+...+q_4^2)$.

\vskip1truecm
The $\lambda \Phi^4$ model in less than four dimensions is superrenormalizable.
The
only divergent graph is that of one vertex with two external legs and a loop.
We can get rid of this divergence simply by renormalizing the mass, and
therefore it is not necessary to renormalize $\lambda$; we can keep it fixed
all the time as we make $a$ go to 0 (equivalently, $\Lambda \to \infty$).
Since $\lambda = \lambda_0 \; a^{-2}$, it
implies $\lambda_0 \to 0$ as $a^2$. In our lattice, consequently, in the
continuum limit $u_4^*=\lambda_0^* = 0$ : we are considering the gaussian fixed
point. Our RGT will evolve in its attraction domain.

   We follow the Renormalization Group scheme; in order
to permit a continuous evolution in the parameter space, we allow integrations
of the variable $p$ between ${ 1 \over s}$ and $1$.   Linearising the resulting
equations near the critical point, we obtain \cite{Rajaraman}

\begin{equation}
\hat \lambda_0 = s^{4-d} \lambda_0;   \qquad  \hat r_0 = s^2
\{ r_0 + {{\lambda_0} \over {4 \pi}} \log s\},
\label{RGT}
\end{equation}
where $\hat \lambda_0$ and $\hat r_0$ are the transforms of
$\lambda_0$ and $r_0$.

   These expressions have a limited region of validity: for big values of
$\lambda_0, r_0$, the linear approximation is not valid.
On the other hand, for
small values of the parameters, we are very near the gaussian point, the
correlation length becomes very large, and if it is of the order of the
lattice size, finite size
effects mask our results. We refer to the intermediate region where the
continuum is reproduced as the scaling region.

   We remark that the fact of staying in the basin of attraction of a Gaussian
fixed point does not prevent at all the possibility of spontaneous symmetry
breaking. Given one point ($\lambda_0, r_0$) in the parameter space, the
Renormalization Group Trajectory starting from it cannot cross the transition
line between the $\langle \Phi^2 \rangle = 0$ and
$\langle \Phi^2 \rangle \not = 0$ phases; it remains in the phase to which the
initial point belongs. Thus, if we start in the symmetry broken phase, the
continuum limit of our theory presents symmetry breaking.

\section{Computation of the soliton and fundamental boson masses}
\subsection{Fundamental boson}
   In order to calculate the mass of the fundamental boson $m_{\rho}$ we use
the connected correlation function between the Higgs fields,
$\langle  \Phi (\vec x,0) \Phi (\vec x,t)\rangle$. In order to avoid
contributions from states with non-zero momenta, we integrate on $\vec x$
and consider \cite{Fox}
\begin{equation}
 C_{\phi}(t) = \langle \phi(t) \phi(0) \rangle  ,
\end{equation}
where $\phi(t) = \int d \vec x \Phi (\vec x, t) .$

   For large $t$, and if the correlation length is different from zero,
$C_{\phi} (t)$ behaves, on an infinite lattice, as
$C_{\phi} (t) \sim \exp (- m t)$.

   We consider periodic boundary conditions in our lattice,L being its extent.
Consequently, the point $n$ is equivalent to $n + L$. Given two points at a
distance $t$, there are two possible paths connecting them: one of length $t$
and the other, resulting from the boundary conditions,  the length of which is
$L - t$.
Thus the mass is given by
\begin{equation}
\langle \phi(t) \phi(0) \rangle \simeq e^{-mt}+e^{-m (L-t)}
\end{equation}
so that
\begin{equation}
{  C_{\phi} (n+1) \over C_{\phi} (n)} =
{{\cosh [m_{\rho}(n+1-{L \over 2})]} \over {\cosh [m_{\rho}(n+-{L \over 2})]}}
\;,
\label{mrho}
\end{equation}
where {\it a} is the time spacing of the lattice. We can solve (\ref{mrho})
and obtain a series of values of $m_{\rho}(n)$ depending on $n$.
For small $n$ they have contributions from large mass states, and for large $n$
the signal is small; there is an intermediate region of $n$ where $m_\rho(n)$
is nearly constant.
We take it as the actual value of the mass.

\subsection{Soliton mass}
   Kadanoff \cite{Kadanoff} introduces the correlation function between two
points in the dual space $R_1, R_2$,
$ \langle \mu_{R_1} \mu_{R_2}  \rangle $ in the following
way: we start from a lagrangian $S= \sum_{n,\mu} J_{n,\mu} \Phi_n \Phi_{n+\mu}
+ \sum_n O(\Phi_n)$; we draw a path in the dual space connecting the dual
points
$R_1$ and $R_2$, and change the sign of the coupling constants J's placed on
the links crossed by our path. We have
\begin{equation}
\langle \mu_{R_1} \mu_{R_2}  \rangle =
{1 \over Z} \sum_{[\Phi]} \exp \{ -\sum_{n,\mu}{}^* J_{n,\mu} \Phi_n
\Phi_{n,\mu} +\sum_{n,\mu}{}' J_{n,\mu}\Phi_n \Phi_{n,\mu}\},
\end{equation}
where
$\sum_{[\Phi]}$ runs over all the configurations of the field,
$\sum_{n,\mu}^*$ takes into account the links with their signs changed and
$\sum_{n,\mu}'$ refers to the rest of the links.
Equivalently,
\begin{equation}
\langle \mu_{R_1} \mu_{R_2}  \rangle = {1 \over Z} \sum_{[\Phi]} \exp \{ S -2
\cdot \sum_{n,\mu}{}^* J_{n,\mu} \Phi_n \Phi_{n,\mu} \} =
\langle   exp \{ -2 \cdot \sum_{n,\mu}{}^* J_{n,\mu} \Phi_n \Phi_{n,\mu} \}
\rangle \;,
\end{equation}
where $S$ is the original action.

 We can alternatively express the correlation function in terms of a ``twisted
action'' $S_t = S -2 \cdot \sum_{n,\mu}^* J_{n,\mu} \Phi_n \Phi_{n,\mu}$,
with it corresponding partition function $Z_t = \sum_{[\Phi]} \exp\{-S_t\}$,
that is to say,
\begin{equation}
\langle \mu_{R_1} \mu_{R_2}  \rangle = { Z_t \over Z}.
\label{Zoverz}
\end{equation}
  In our case, with a lagrangian given by (\ref{discrete}), we can express it
in a similar way, depending on link variables $J_{n,\mu}$,
after making a change of variable  $\Phi \to \sqrt{J} \xi$. Now,  $J_{n,\mu} =
{\rm constant}
= J
\; > 0$. This causes the appearance of a twist:  the fields placed in the
points $n,n+\mu$ connected by a link where $J_{n,\mu}$ has changed to -
$J_{n,\mu}$ tend to change their signs: we have given rise to a topological
excitation, with a non-zero topological charge $\Phi(x=\infty)
-\Phi(x=-\infty)$.

   We now define
\begin{equation}
C_{\mu}(t) = \langle \mu(n,t) \mu(n + \tau ) \rangle ,
\end{equation}
where our
path in the dual lattice will be the minimum length path connecting them, i.e.
straight vertical  lines.

   The topological excitation (with non-zero projection on the soliton sector)
appears at the time $t$, and annihilates at $t + \tau$. Thus, we expect an
exponential behaviour, similar to that of  $C_{\phi} (t)$. In this case, one of
the two paths connecting the points $R_1$ and $R_2$ has a much bigger
contribution to $C_{\mu}$: that which crosses the dislocation. In fact,
we have observed  a clear
exponential decay, $\langle  \mu_ {(\vec n, t)} \mu_{(\vec n, t+\tau)} \rangle
\sim  \exp \{- C \tau \}$,  and therefore we obtain the soliton mass as
\begin{equation}
m_{soliton} =- \log {C_{\mu} (t) \over C_{\mu} (t+1) }
\label{msoliton}
\label{msol} \;.
\end{equation}

   When we study  $\langle  e^{-2 \cdot \sum_{n,\mu}^* \Phi_n
\Phi_{n,\mu}}\rangle$, we must consider the risks of our method: we study
a strongly non-local quantity, which is seriously affected by the finite size
of our lattice. Besides, the use of an exponential function implies a
magnification of errors.

 In principle, we do not know to what point these effects will spoil our
results.
In order to control these risks, we look for an alternative way of calculating
the soliton mass from local non-exponential variables. Following Groeneveld et
al.  \cite{Groeneveld}, we introduce a local parameter $\Omega(\beta)$, which
accounts for the energy response to the appearance of the twist.

   First, it will be useful to change the variables the action depends on.
 We note that, making the following change of variable,
$\Phi \to \xi= {{\Phi} \over {\sqrt \beta_0}}$ with
$\beta_0={1 \over \lambda_0}$, we
obtain
\begin{eqnarray*}
Z(r_0,\lambda_0) \equiv \hat Z(r_0,\beta_0) =
\end{eqnarray*}
\begin{equation}
\beta_0^{-{V \over 2}} \int_{- \infty}^{+
\infty}  (\Pi_n d\xi_n) exp \{ - \beta_0 \,
[  - \sum_{n,\mu} \xi_n \xi_{n,\mu} + \sum_n
\{  (d+{{r_0} \over 2}) \xi_n^2   +  { 1 \over {4!}} \xi_n^4\}]
\,\} \;,
\end{equation}

where $V = L^d$ is the volume of the system. The $\beta_0$ derivative of
$Z(r_0,\lambda_0)$ is
\begin{equation}
{{\partial \hat Z(r_0,\beta_0)} \over {\partial \beta_0}} = - {V \over 2}{{\hat
Z(r_0,\beta_0)} \over \beta_0}  - {1 \over \beta_0} Z \langle
S[r_0,\lambda_0]\rangle,
\end{equation}
where $S(r_0,\lambda_0)$ is the action resulting from the integration of our
lagrangian (\ref{discrete}).

  If we impose antiperiodic boundary conditions in the spatial direction, we
introduce a twist the length of which is the temporal dimension of the lattice,
$T$. We can define the ``twisted'' partition function corresponding to  this
twist, $Z_t(r,\lambda) \equiv \hat Z_t(r,\beta)$.
 Keeping in mind (\ref{Zoverz}) we can now calculate the soliton mass as
\begin{eqnarray}
m_{\rm sol} =  - {1 \over T} \log \langle \mu(\vec n,T) \mu (\vec n,0 )\rangle
=
- {1 \over T} \log {{Z_t(r_0,\lambda_0)} \over {Z(r_0,\lambda_0)}} = \nonumber
\\
 - {1 \over T} \log {{\hat Z_t(r_0,\beta_0)} \over {\hat Z(r_0,\beta_0)}} =
\int_{\beta_c}^{\beta_0} \Omega (\beta_0'),
\label{integral}
\end{eqnarray}
where $\beta_c$ is the
value for which our trajectory cuts the transition line between the
$\langle \Phi^2 \rangle = 0$ and $\langle \Phi^2 \rangle \not = 0$ phases, and
\begin{equation}
\Omega (\beta_0)=  - {1 \over T} {\partial \over \partial \beta_0}\log {{\hat
Z_t(r_0,\beta_0)} \over {\hat
Z(r_0,\beta_0)}} = {1 \over \beta_0}{1 \over T} \langle  S_t(r_0,\lambda_0)
\rangle_t -
\langle S(r_0,\lambda_0)\rangle  \;,
\label{omega}
\end{equation}
where $\langle \rangle$ and $\langle \rangle_t$ stand for expectation
values with $Z$ (periodic boundary conditions) and $Z_t$ (twisted or
antiperiodic) respectively. We remark that in the $Z_t$ system
$\langle \Phi \rangle = 0$ in both the symmetric and broken phases.
   The integration in (\ref{integral}) implies defining a trajectory in the
parameter space with $r_0$ fixed, and $\beta$ starting from $\beta_c$.
For higher values of $\beta$, we have no symmetry breaking, and the soliton
mass vanishes.

  We will check the masses  obtained with the exponential function by comparing
them with those resulting from using $\Omega$.

\section{Details of the simulation}

   We have made use of a specially-designed transputer based parallel machine,
RTN, including 64 T-805 processors distributed in 8 boards with 8  each. As an
individual
board calculates one point in the parameter space, we get eight absolutely
independent groups of measurements for every ($\lambda_0$,$r_0$). The error for
every magnitude has been calculated averaging its 8 independent predictions.
We have used an adaptative MC process so as to keep the rate of acceptance
between 40 \% and 60 \%.

   We have simulated different lattice sizes ($ 16^2, 24^2, 48^2$ ), with
(1000, 3000, 7000) iterations of thermalization and (2000, 22500, 30000)
measurements.  Within each transputer,
we have taken (20, 10, 5) decorrelation MC iterations between two
consecutive measurements.We have observed no relevant finite-size effects in
the
local quantities. However, big sizes are needed when computing correlations,
especially those defined by Kadanoff's operator, as a consequence of its
strong non-locality and its exponential form.
As we have mentioned earlier, we can not rely on the small-length
correlations because of the contribution of large mass states; on the other
hand, long distance correlations are seriously affected by the finiteness of
our lattice. Thus, in order to obtain a precise value for the
mass from Kadanoff's  operator we have needed bigger and bigger lattices.

\section {Phase diagram and the scaling region}

   Our model exhibits two phases. Classically, for positive values of $r_0$
the minimum energy configuration is $\Phi = 0$; for negative $r_0$, a
spontaneous symmetry breaking occurs, and the new minima are
$\Phi = \pm \sqrt {{6 |r_0|} \over {\lambda_0} }$.

  When we consider the contributions of all the configurations, each weighted
with $\exp \{ -S_{euc}  \}$, for small negative values of $r_0$ both minima are
very close to each other and are not deep enough to stop fluctuations from
restoring the symmetry.
More negative  values of $r_0$ are necessary to ensure that we are in the
broken phase.
Therefore, in the semiplane with  negative $r_0$ there is a transition line
separating both phases.

   In order to determine the transition line, we choose several values of
$\lambda_0$. For each of them,  we decrease $r_0$ until $\langle \Phi^2
\rangle $ becomes different from zero; in the limit of an infinite volume, its
value passes from zero to a finite non-zero value when crossing the
transition line; in a finite-volume system in the symmetric phase,
$\langle \Phi^2 \rangle \approx 1/\sqrt{V}$,  and what we see is a sharp
rise of $\langle \Phi^2 \rangle $ (technically, in a finite lattice $\langle
\Phi \rangle $ is not a good order parameter because tunnelling between states
with positive and negative values of the field cause it to be equal to
zero all over the parameter space). Another useful quantity as an order
parameter is the soliton mass. When computing $\langle \exp \{ -2J \sum
\Phi_n \Phi_{n+\mu} \} \rangle$, if we are in the $\langle \Phi^2 \rangle  = 0
$ phase, the values of $\Phi_n$ fluctuate around zero, and their sum over the
path vanishes; the expected values appearing in (\ref{msol}) become
independent of the length of the path and equal to $1$ and $m_{sol}$ is zero.
On the other hand, we expect non-zero $m_{sol}$ for the symmetry broken phase,
where the $\sum \Phi_n \Phi_{n+\mu}$ is different form $0$.
We determined the transition line using both parameters ( $\langle \Phi^2
\rangle $ and $m_{sol}$). It is shown in figure 1.

   In the region where
${{|r_0|} \over {\lambda_0}}$ is large enough, we can compare them with mean
field predictions. Theoretically, $|\Phi| =
\sqrt{{{6 |r_0|} \over {\lambda_0}}}$.   In this region
the fluctuations are small, and thus we have
$\langle \exp \{ -2J \sum_L \Phi_n \Phi_{n+\mu} \} \rangle
\sim  \exp \{-2JL|\Phi^2| \}$, where $L$ is the length of the summation path,
and we can consider $\langle \Phi(0) \Phi(r)\rangle \sim |\Phi^2|$
and therefore $m_{sol} = 2J|\Phi^2|$. Our results agree with these predictions.

   Now we pass to determine the scaling region. We keep ourselves in the
$\langle \Phi^2 \rangle \not = 0$ phase, where $m_{\rho}$ and $m_{sol}$ are
different from zero. The reason for this is that, proceeding in this way, the
continuum limit of our theory will correspond to the symmetry broken phase of
the continuum problem, which is the one we are interested in.

   Thus, our next step is finding the region in the parameter space where
equations (\ref{RGT}) are valid.

   Along a RGT we expect to find constant values for the physical meaningful
variables, such as the correlation length $\xi$, the physical masses M...
In the lattice we work with adimensional quantities depending on the point
of the trajectory ($\xi_0(r_0,\lambda_0),m(r_0,\lambda_0)...$), related to
the physical ones by  $M= {m \over a}, \xi = a \cdot \xi_0...$    Consequently,
although
our lattice-defined quantities vary, the ratios
$m_{\rho}   / m_{sol} \;,  \;m_{\rho}^2  / \lambda_0 \;,  \;
m_{soliton}^2  / \lambda_0$,  which are equal to the physical expressions
$M_{\rho}  / M_{sol} \;,  \; M_{\rho}^2  / \lambda\;,  \;
M_{sol}^2  / \lambda$, remain constant along these trajectories.

   We start from different points in the parameter space $(r_0,\lambda_0)$ near
to
the critical point $(-0.2< r_0 < 0,  \lambda_0 ={\rm fixed} = 0.1)$. Iterating
Renormalization Group transformations (eq. (\ref{RGT}),where  we choose
s=1.08), we get further and further away from the origin (and thus from the
continuum limit), drawing a series of
trajectories. Along each one of them, we calculate the previously defined
ratios in the different points obtained by the transformations. For every
trajectory,  we find a segment where these quantities remain approximately
constant;   the union of all the segments gives us the scaling region.

   Initially, we follow curves near the transition line separating the
$\langle \Phi^2 \rangle =0$ and $\langle \Phi^2 \rangle \not = 0$ phases. As
we move away from it,  we find that the length of the segment reduces. This is
clear, because we need a large lattice correlation length in order to
reproduce the continuum limit, and the region close to the line transition
is appropriate to that. Far from this line the correlation length is small,
and the discretization is important.

   Finally, we choose a curve near that line, with the initial values \\
($r_0$= -0.105,   $\lambda_0$= 0.25) (see figure 1).
At this point we can illustrate our comments about the difficulties derived
from
the use of Kadanoff's operator. In figure 2 we represent the correlations of
the fields $\Phi$ and $\mu$ for some points of this trajectory.  As expected,
when we get near the gaussian point, the correlation length increases, the
mass is lower and
the correlation decreases more and more slowly.  For small enough values of
the parameters, the correlation function $\langle \mu_N \mu_{N+n} \rangle$,
for distances of the
order of the length of the lattice, is not compatible with zero. ($n$ is the
distance in units of the lattice spacing). Thus, we must
be very careful when we calculate masses in this region.

   There is another reason that makes it desirable to work with big lattices.
Our
method for calculating the masses consists  basically on finding a certain
correlation function, and fitting it to an exponential, or to an hyperbolic
cosine. We expect this fit to be reasonably good for a set of intermediate
values
of $n$.   When $n$ approaches the length of the lattice -in our case, half
this length, because of the periodic boundary conditions-, the fit is not
possible any longer. The bigger our lattice is, the longer this well-fitting
segment becomes, and we have more points to fit our theoretically predited
behaviour, and so calculate the mass with higher precision. This is clearly
shown in figure 3: in the plot at left we draw the logarithm of the
correlation function $\langle \mu_N \mu_{N+n} \rangle$ for $24^2$ and
$48^2$ lattices, in a region far from the gaussian point. In the small
lattice, when $n \sim 9$, the fitting to a straight line is no longer possible,
while in
the big one we can  still include some more points and get a good fit to a
straight line. In the small picture at
right, the parameters are $\lambda_0= 0.25$, $r_0 = -0.105$; we have seen in
figure 2 that, for these values, the correlation length is comparable with
the lattice length, and we expect serious corrections. In fact, for the
small lattice, the agreement region is smaller.

   In figure 2 we see that the function $\langle \Phi_N \Phi_{N+n} \rangle$
is much smaller than the $\mu$ correlation, and we expect that the values
obtained
for the boson mass are not so strongly affected by the size of the lattice.
Our results confirm this prediction.

   Now we can estimate the scaling region. From figure 4, we see that it begins
at $\lambda_0 \sim 2$, the value from which $m_{\rho} / m_{\rm sol}$ can be
considered as a constant. The upper boundary of this region can be more
clearly inferred from figure 5.   We expect $m / \sqrt{\lambda_0}$ to be
constant or, equivalently, a linear behaviour of $m$ with $\sqrt{\lambda_0}$,
$m$ tending to zero as $\sqrt{\lambda_0}$ does.  Thus, for the RGT starting
from the initial values $\lambda_0 = 0.25$, $r_0 = -1.05$, the scaling region
corresponds to the interval  $ 1.5 < \lambda_0 < 14 $

\section {Results}

\subsection {Results from the operators}

   First of all, we want to check that what we call soliton mass, calculated
using Kadanoff's operator, really behaves as a mass. We will
compare its evolution under the Renormalization Group equations to that
of the fundamental boson mass. On the other hand, we compare its value to
previous theoretical predictions.

   A quantity $\Theta_{phys}$ with dimension [$\Theta$] is related to its
equivalent in the lattice  $\Theta_{latt}$ by $\Theta_{phys}  \sim
a^{[\Theta]} \Theta_{latt}$. As we are working in less than four dimensions,
we can keep $\lambda$ fixed, and $\lambda_0 \sim  \lambda \cdot a^2$ tends to
0 as $a^2$ when we approach  the continuum limit (gaussian fixed point)
along a RGT. We also expect finite values for $M_{\rho}, M_{sol}$, so our
value of the masses in the lattice $m \sim M \cdot a$ tends to 0 as $a$.

 From that we deduce, as we approach  the continuous limit, $m_{sol}
\sim  \sqrt{\lambda_0}$. Figure 5 shows  that, within
the limit of the scaling region, that is the behaviour of $m_{\rho}$ and
$m_{sol}$.

   In a system with only one relevant direction, the renormalized trajectory
coincides with that direction. In our case we have a twice-unstable point, and
there is a continuous family of renormalized trajectories leaving it, each of
them corresponding to a different continuum theory. In the previous section we
have chosen one of those trajectories; once we give an arbitrary value of
$\lambda$, $M_{\rho}$ and $M_{sol}$ can be calculated as $M = {m \over \sqrt
\lambda_0} \sqrt \lambda$. In our RGT, $M_{\rho}=  0.453 \sqrt \lambda$ and
$M_{sol}= 0.356 \sqrt \lambda$.

   Next we study the evolution of $m_{sol}$ when, starting from a point on the
transition line, we move further and further into the symmetry broken phase.

   First, let us summarise some qualitative basic ideas. When we quantise the
classical absolute minimum,
we obtain the vacuum of the quantum theory; quantisation around the soliton
gives us the soliton sector. If this local minimum is broad (which, in our
case,
corresponds to a point near the transition line between $\langle \Phi^2 \rangle
\not = 0 $  and  $\langle \Phi^2 \rangle = 0 $), a great number of
configurations different from the classical solution will contribute to
the value of any observable. But as we move away from this zone, the potential
well gets deeper and deeper, and a moment comes when we have contributions only
from the minimum and  configurations very close to it; we are recovering
the classical solution.

   In order to explore these ideas, we trace a trajectory in ($\lambda_0,r_0$)
space
fixing $r_0$= -2.2 and letting $\lambda_0$ move from the vicinity of the
transition
line ($\lambda_0 \simeq  12$) deeper and deeper into the
$\langle \Phi^2 \rangle \not = 0$ phase.   This path cuts a different RGT in
each of its points (with different physical masses for a value of $\lambda$).

   To calculate the classical continuous limit  for the soliton mass, we
make use of the fact that, in a continuous euclidean space, a soliton of the
$\lambda \Phi^4$ classical theory in 1+1 dimensions propagating with a velocity
$v$ is given by \cite{Rajaraman}
\begin{equation}
\Phi(x,t) = \sqrt{{6 |r|} \over {\lambda}}  \;\tanh \,\lbrack  {|r|^{1/2}
\over {\sqrt 2}}
( {{x - vt} \over {\sqrt{(1-v^2)}}})
\rbrack.
\end{equation}

   The energy density is $\epsilon(x)$, the expression of which coincides with
the euclidean lagrangian. The classical absolute minimum is $\Phi = \pm \sqrt
{{6 |r|} \over {\lambda}}$. At a classical level, we can define the soliton
mass as
\begin{equation}
M_{class}= E - E_{min}= 4 \sqrt{2} {{|r|^{3 \over 2}} \over {\lambda}} \;,
\label{mclas}
\end{equation}
$E_{min}$ being the energy of the absolute minimum, and $E$ that of the
soliton solution.

  When we quantise the soliton, in the weak coupling approxi\-mation
(${{\hbar \lambda} \over r} << 1$),  the mass obtained is, up to
an order $\Theta (\hbar \lambda / |r| )$ \cite{Rajaraman}

\begin{equation}
M_{quantum} = M_{class} + \hbar \sqrt{r} ( {1 \over 6} \sqrt{3 \over 2} -
{3 \over {\pi}} \sqrt 2) \;.
\label{quantum}
\end {equation}
We have taken all along $\hbar$ equal to 1.

   Comparing our results for $m_{sol}$ to those predicted by
expression (\ref{quantum}) (see, e.g., figure 7), we see a clear linear
behaviour with $\beta_0 = 1/ \lambda_0$.
That behaviour is intermediate between the
O(0) and O($\hbar$) theoretical predictions, closer to this second one.
  That displacement with respect the  O($\hbar$) prediction may be
attributed to the
contribution of higher orders in $\hbar \lambda_0 / |r_0|$.

\subsection{Results from twisted system}

   The use of twists is known to be, in a computer simulated theory, a good
help for studying the phase structure. Our main motivation for introducing it
is to calculate $m_{sol}$ in an alternative way to the use of Kadanoff's
operator, so avoiding its risks, already mentioned in section 2.

   We impose antiperiodic boundary conditions in the spatial coordinate. In
this
way, we introduce a twist in the lattice that lasts from t=0 to t=T. We
evaluate
the expected value of the action under these conditions,
$\langle S_t \rangle_t$, following the notation introduced in section 2.
In the $\langle \Phi^2 \rangle = 0$ phase, because of its $\pm \Phi$ symmetry,
changing the signs of some J's does not cause the expected value of the
action to change, and $\langle S \rangle = \langle S_t \rangle_t$. That means
$m_{sol}=0$,
or we can also see the vacuum as a ``soliton condensate''.
However, in the  $\langle \Phi^2 \rangle \not = 0$ phase things are not so any
longer. By inducing the twist, we favour the appearance of a soliton
propagating
through time.$\delta S = \langle S_t \rangle_t - \langle S \rangle$ is the
energetic
response of the lattice
to the introduction of the twist;  it must be related to
the euclidean energy of the soliton, as we will see.

   As we did in the previous section, we follow the path fixing the value
of $r_0$ = -2.2.
$\delta S$ must increase its value from 0 near the transition line to the
classically predicted one. Again, we plot $\delta S$ against
$|r_0|^3 / \lambda_0^2$. In figure 6, we can see that it grows up steeply
and soon stabilizes at the classical value (\ref{mclas}):
$|\delta S|^2 / \lambda_0  = 32
|r_0|^3  / \lambda_0^3$. Instead of $\lambda_0$ we have drawn $\beta_0 = 1 /
\lambda_0$.

   Intuitively, there is a relationship between $\delta S $ and the soliton
mass, which is
its minimum energy level. In the classical limit, $\delta S / T$ is
equal to $m_{sol}$. In figure 6, we see that, as we increase $\beta_0$, both
values tend to
coincide. In general, the expression relating both quantities can be obtained
if we keep in mind (\ref{integral}) and
(\ref{omega})
\begin{equation}
m_{sol} = {1 \over T} \int_{\beta_c}^{\beta}  {{\delta S(\beta')} \over
 {\beta'}} \;.
\label{connect}
\end{equation}

 From fig. 6, we see that
$\delta S / T $ suffers a sharp increase around $\beta_0 = 0.0826$,
($\lambda_0 = 12.1$) indicating that
we have crossed the transition line. We deduce $\beta_c = 0.0804 \pm 0.0022$
$\lambda_c = 12.45 \pm 0.35$. In order to avoid
errors coming from the estimate of $\beta_c $ and $\delta S$ in the vicinity
of that line, we take, instead of (\ref{connect})
\begin{equation}
m_{sol}(\beta) = m_{sol}(\beta_i) + {1 \over T} \int_{\beta_i}^{\beta}
{{\delta S(\beta')} \over {\beta'}} \;,
\end{equation}
where we have taken $\beta_i=0.0826$, the first value of $\beta$ for which
$\delta S$
is clearly non-zero. We fit our values of $m_{sol}$ to a straight line, and
take
$m_{sol}(\beta_i)$ to be the height of the line at $\beta_i$.

 We compare the results obtained applying (\ref{integral}) with those from
Kada\-noff's operator (see figure 7). Both values coincide with a precision
up to 3 \% in the least favourable point.

\section{Conclusions}

 We have studied topological excitations in the $(\lambda \phi^4)_{1+1}$ model
on the lattice using a ``disorder parameter'', from the decay of which we
can compute the soliton mass.   The results  obtained
have a well defined continuum limit, which we have computed with
the renormalization group equations.   We have paid special attention to
make sure that
we are working in the scaling region, where
physical quantities are unchanged along the RGT.

   We have also computed the soliton mass from the difference of the vacuum
energy between the twisted an untwisted systems, where quantities are local,
and we found this result agrees with te previous one.

Now we would like to compare both methods: the first has the disadvantages that
it implies the calculus of operators which are strongly non-local and
exponential, and thus the method is very sensitive to finite-size effects and
systematic errors. A lot of statistics is necessary to obtain results within a
reasonable margin of error. On the other hand, the method using twisted systems
decreases considerably the computation time required. The statistic errors are
small, and thus we conclude that imposing twisted conditions is a very
satisfactory alternative in order to calculate the soliton mass. However,
twisted conditions modify the vacuum of the theory, and make  the calculation
of other masses to which to compare the results (such as, e.g, the fundamental
boson mass) impossible.

   We have compared our non-perturbative result for the mass of the topological
excitations in the continuum limit with the theoretical perturbative result up
to first order. Our results show a
systematic lineal raise, which may be due to
higher order corrections not considered in the perturbative prediction.

The inclusion of fermions with a Yukawa coupling to the scalar fields
is a very interesting future work. In this case we have a three parameter
space and a very  rich model. The problem is simplified by the fact that
bosonization is possible, and so
Monte Carlo simulation is simpler than when fermions are considered
directly. Therefore,  it must be also possible to compute the soliton
mass and the continuum limit.

\section{Acknowledgements}

   We want to thank A. Cruz, L.A. Fern\'andez, P. Di Giudice and A. Grillo
for their kind suggestions, and the RTN group for the use of the RTN machine.

\newpage

{\large Figure captions}

\begin{itemize}

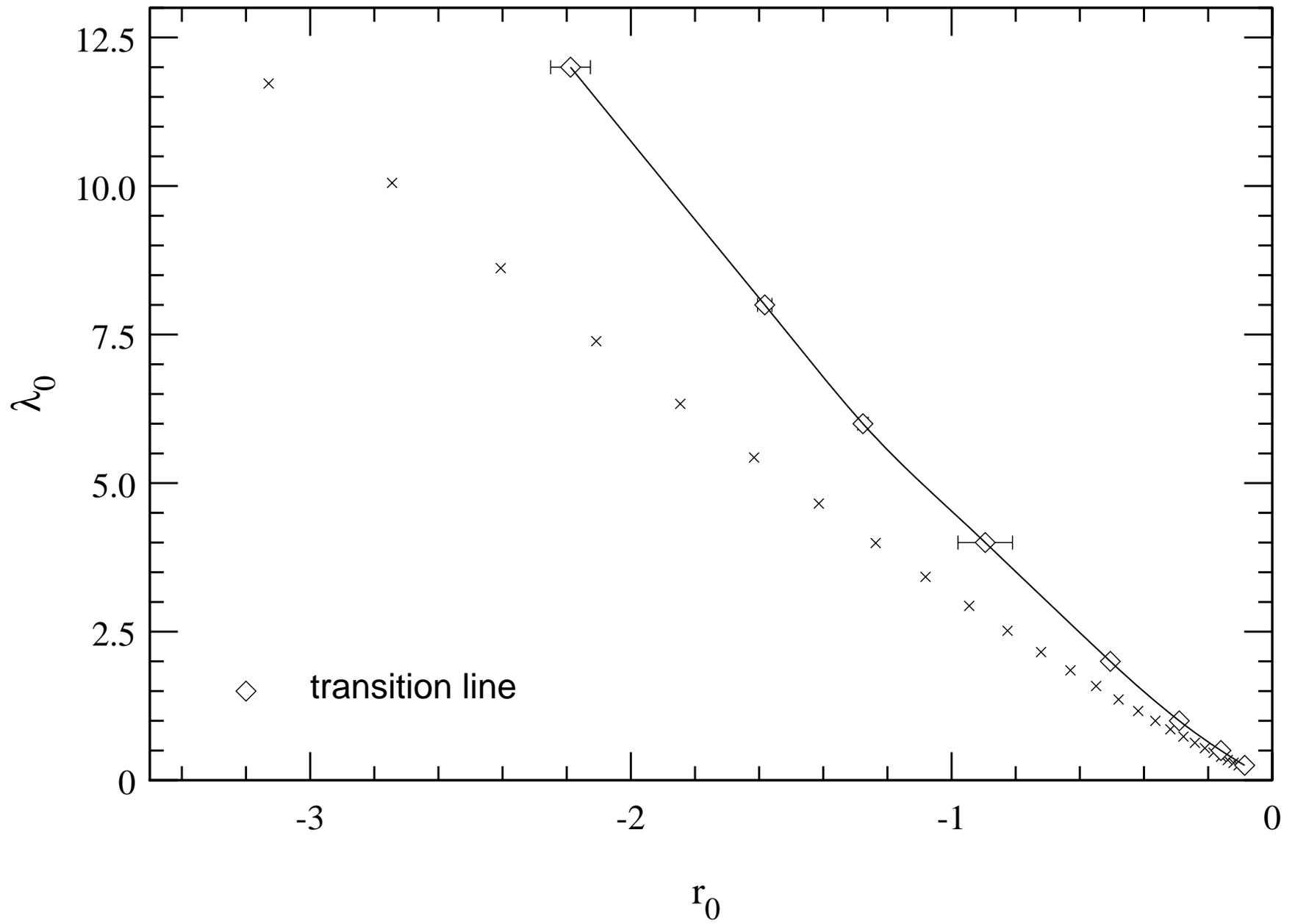
\item[Fig. 1] The Renormalization Group Trajectory  followed
starting from the initial values $r_0= -0.105$, $\lambda_0 = 0.25$ is shown.
The transition line between the $\langle \Phi^2 \rangle = 0$
and $\langle \Phi^2 \rangle \not = 0$ phases is drawn.

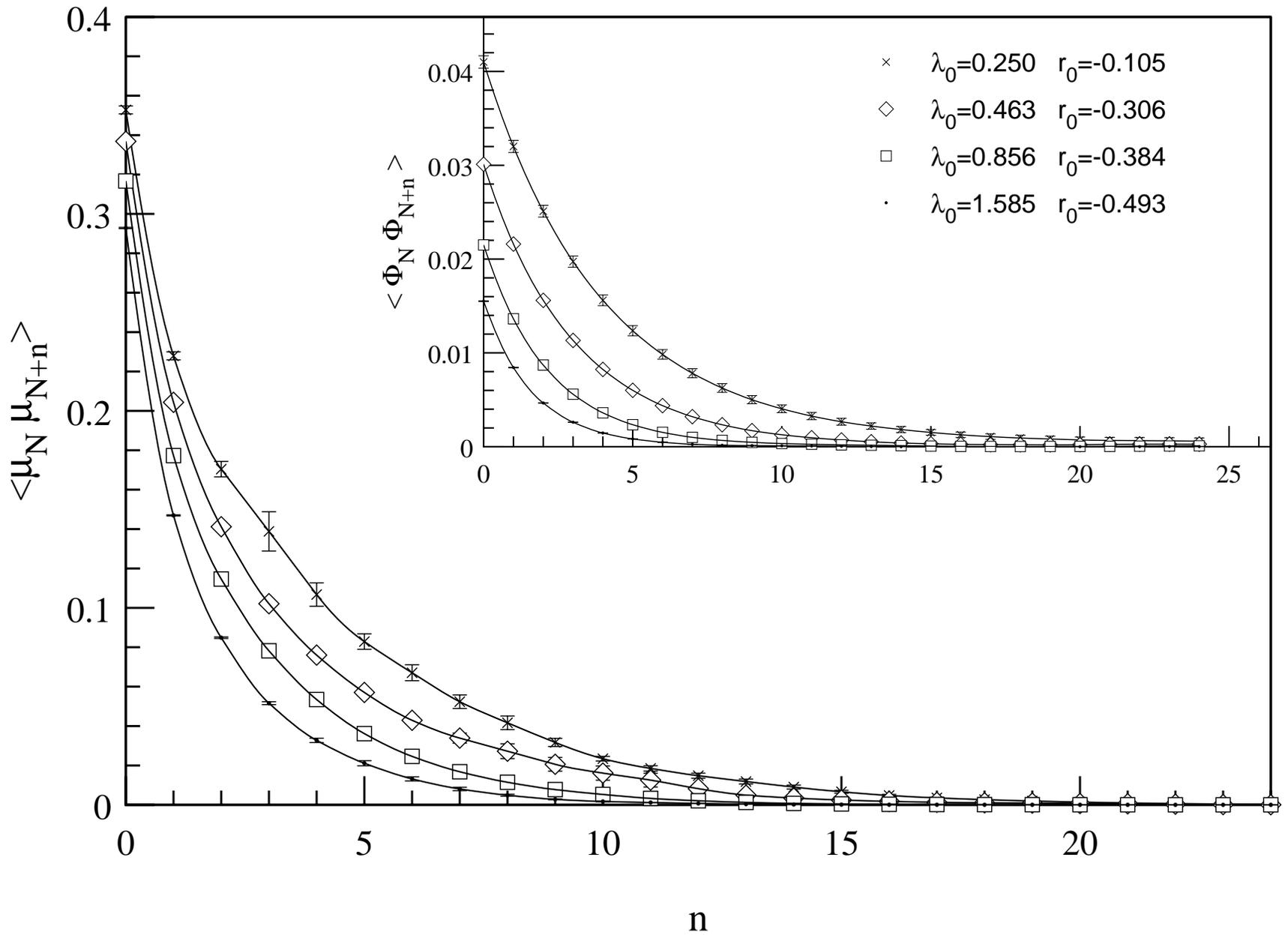
\item[Fig. 2] Decay of the correlation functions $\langle \mu \mu \rangle $
and $\langle \Phi \Phi \rangle $ in a $48^2$ lattice. While the latter decays
quickly to 0,   $\langle \mu \mu \rangle $ is non-compatible with 0 for
large distances if we are near the gaussian point. Thus, we expect the
value of the mass we calculate from it to improve as we increase the
lattice length.

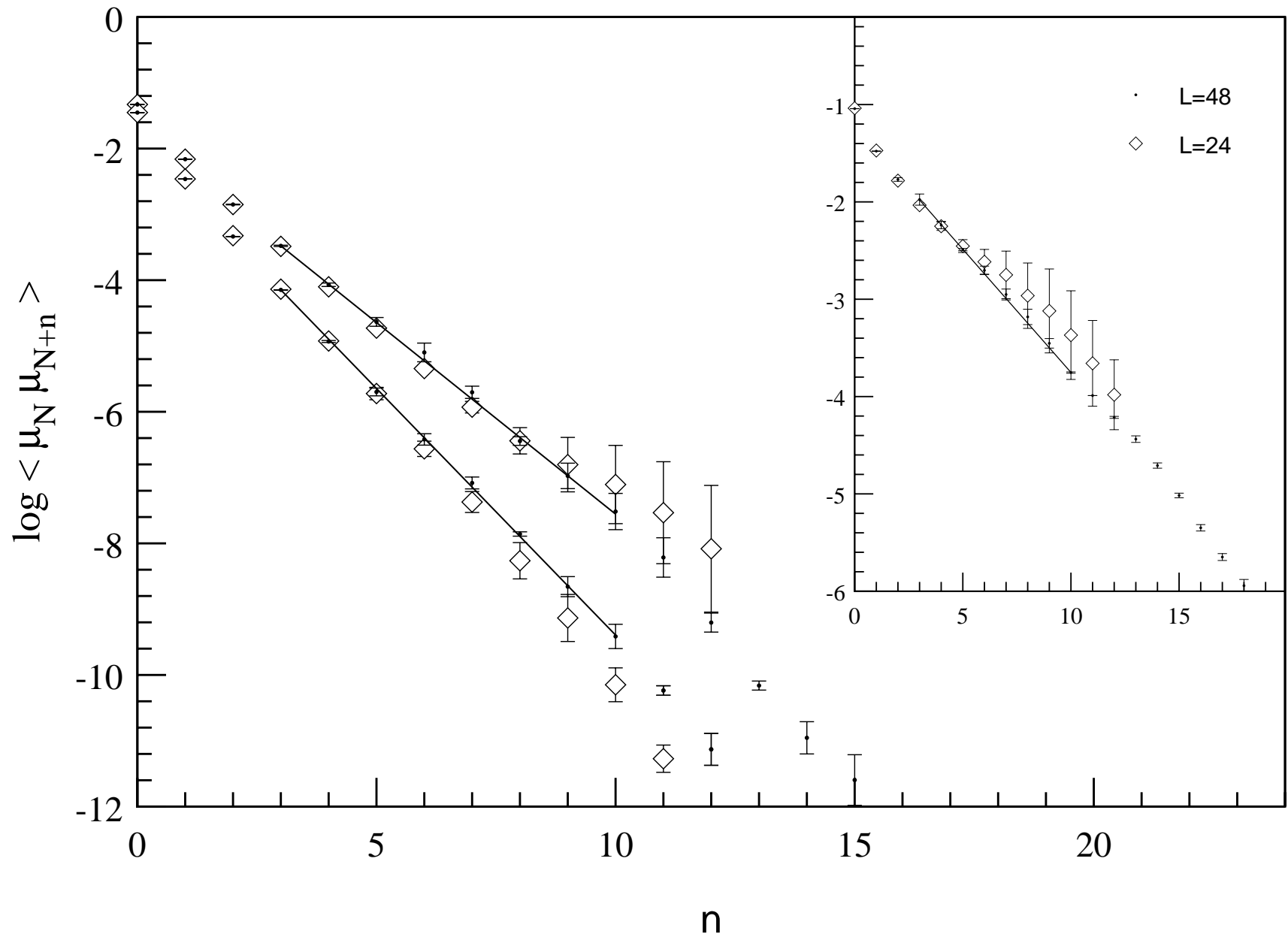
\item[Fig. 3] Logarithm of the correlation function
$\langle \mu \mu \rangle $ for three different points in the parameter space:
for the small window at right, $\lambda_0 = 0.25 $, $r_0= -0.105$. The two
other points are
 $\lambda_0 = 2.934 $, $r_0= -0.945 $ and $\lambda_0 = 5.431 $, $r_0= -1.616$.
Both are inside the scaling region.
The latter is the one with
the largest slope (the biggest mass).

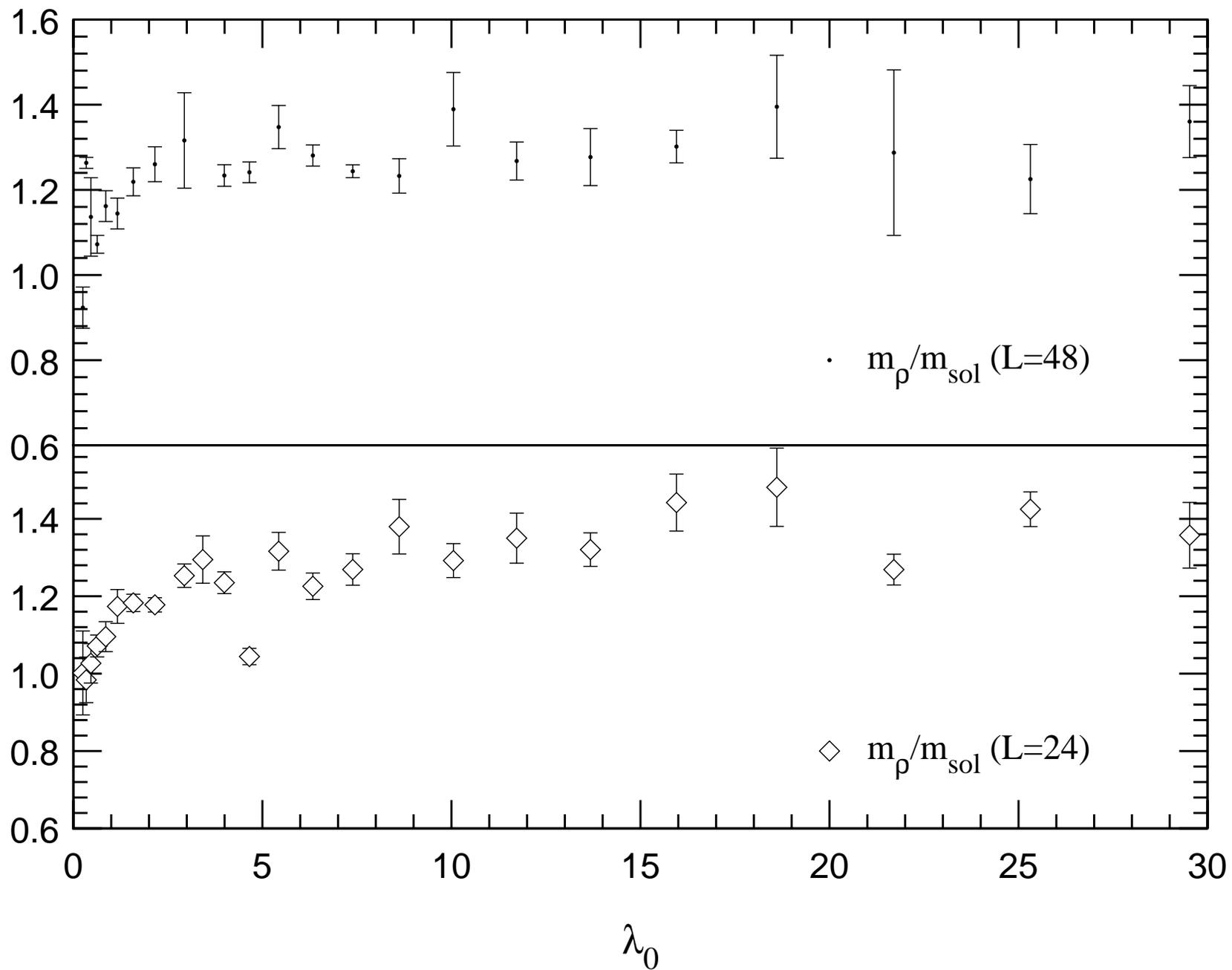
\item[Fig. 4] Evolution of the ratio $m_{\rho} / m_{sol}$
along the RGT previously drawn in figure 1. From the results in the $48^2$
lattice,
we see that the ratio becomes constant from $\lambda_0 \sim 1.5$,
a fact which is not apparent in the $24^2$ lattice. This gives us a
lower limit for the section of our RGT inside the scaling region.

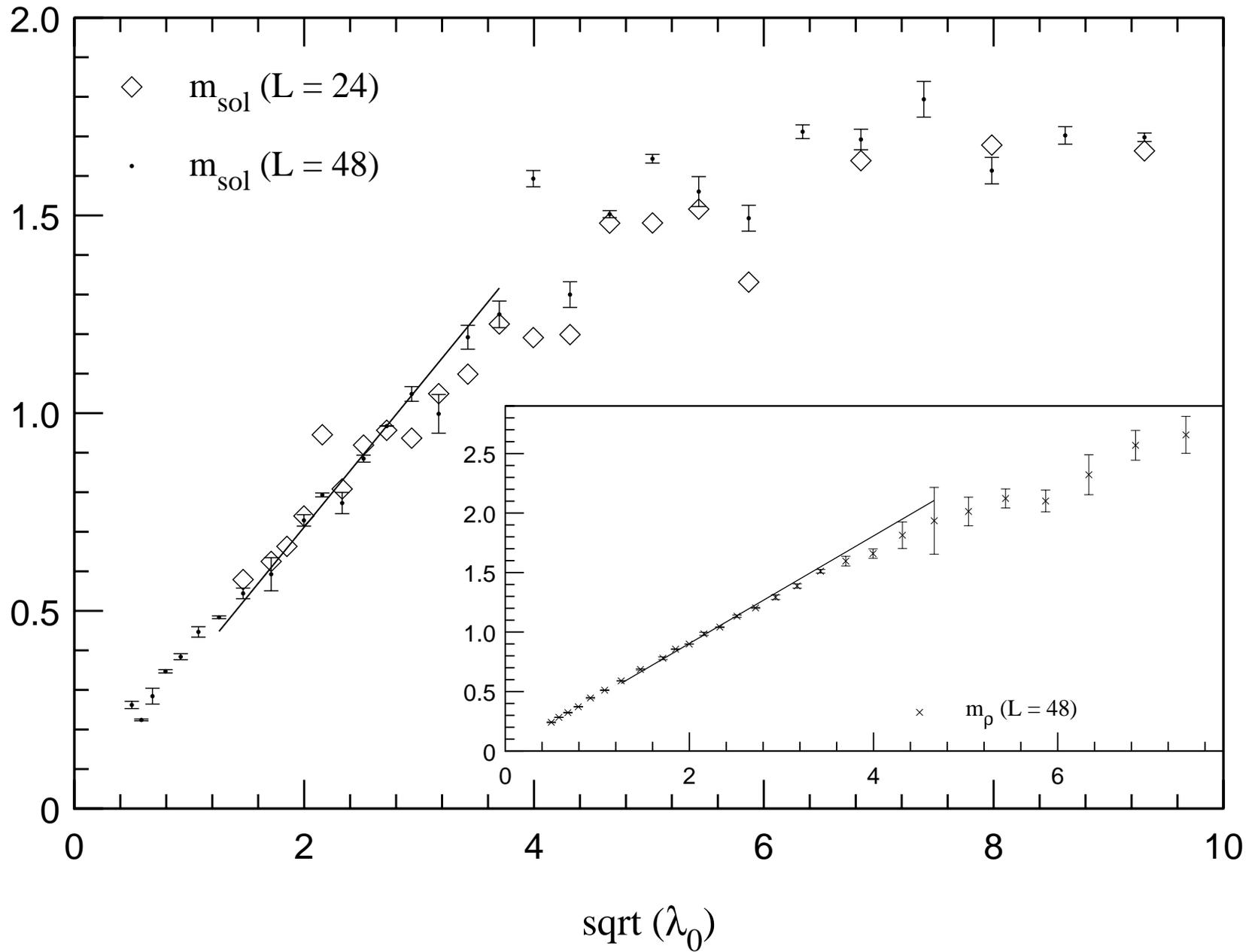
\item[Fig. 5] ${m_{\rho}}$ and
${m_{sol}}$ versus $\sqrt{\lambda_0}$  along the RGT, for $24^2$ and
$48^2$ lattices. We determine the
scaling region by selecting the points which give a reasonable fit to a
function  $y= a \cdot x $. In this way, we find the upper
limit of this region,
given by $\lambda_0 \sim 14$. For $m_{\rho}$  only  the results from
the $48^2$ lattice are drawn because they coincide with those from the small
lattice.

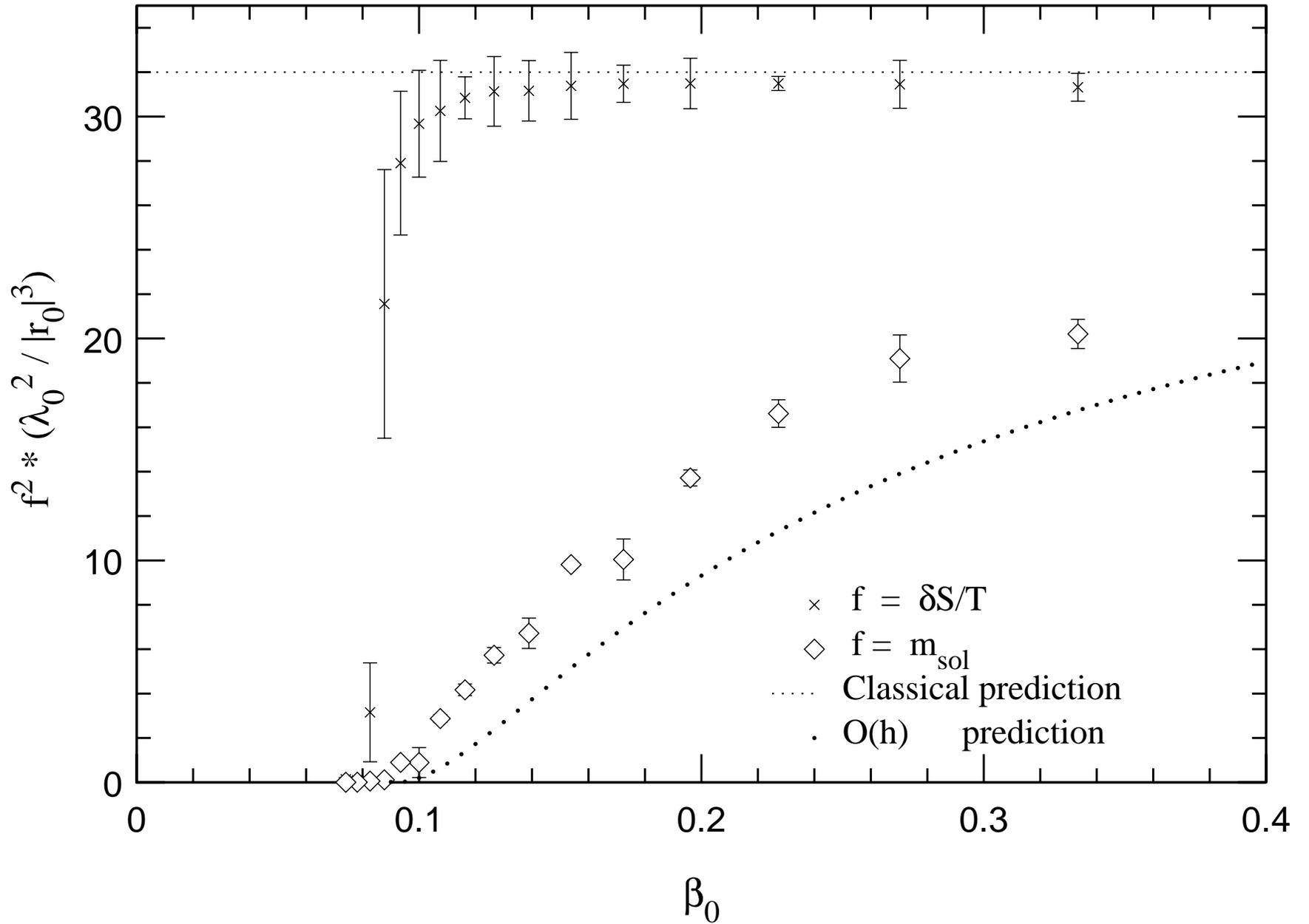
\item[Fig. 6] In a $48\times 48$ lattice, along the $r_0 = -2.2$
vertical path, the evolution of the quantities $ (\delta S /  T )
(\lambda_0^2 / |r_0|^3) $ and
$m_{sol}^2  (\lambda_0^2 / |r_0|^3)$ versus $\beta_0 = {1 \over \lambda_0}$
is shown.
The classical value for both ratios is 32. The results are compared to
those for the  perturbative calculation up to an order
$O ( \lambda_0  / |r_0|)$. While all these results coincide in
the limit $ (\lambda_0  / |r_0| ) \to 0$, $(\delta S / T)$ soon
stabilizes at the classical value, while $m_{sol}$ keeps closer to the
first-order
calculation.

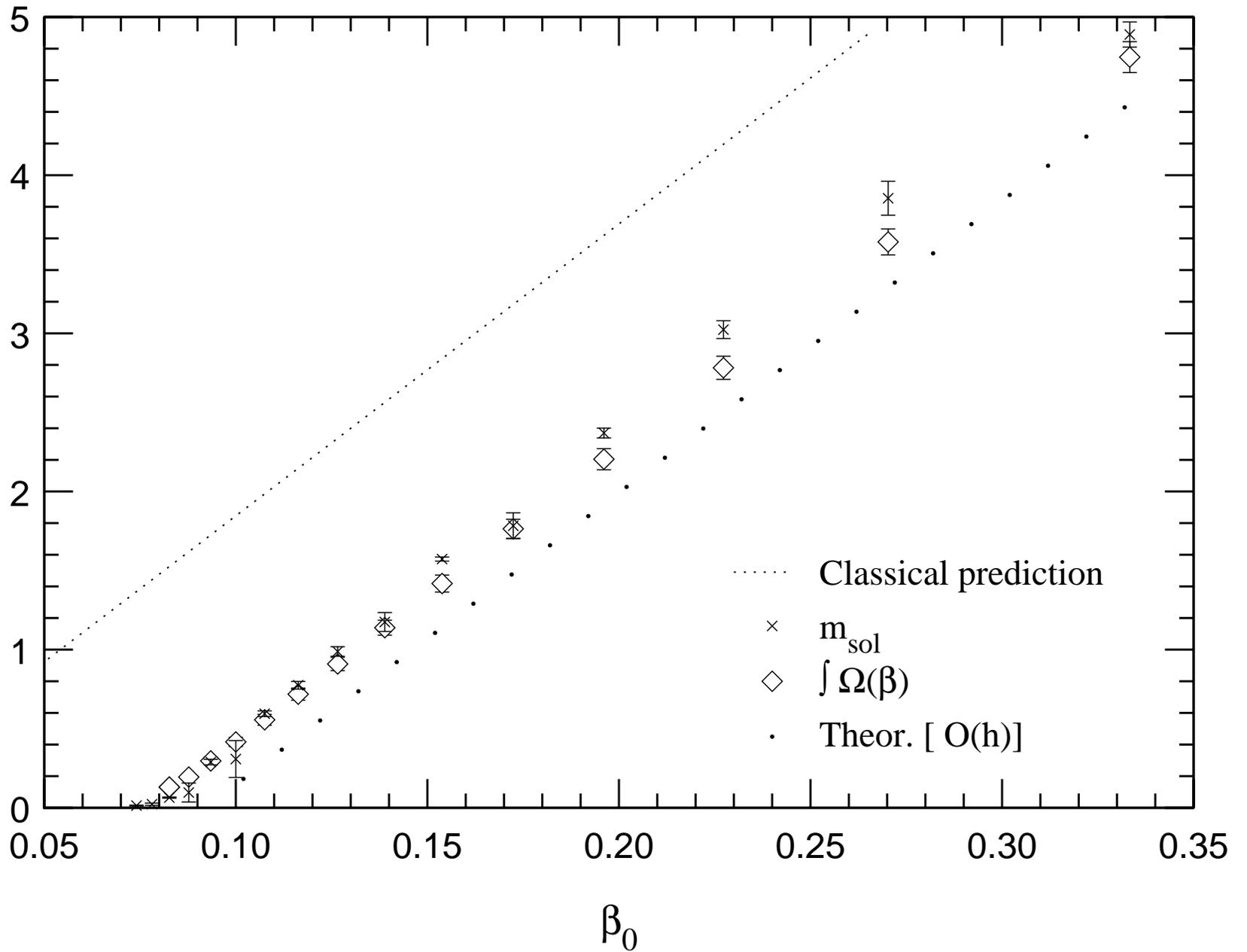
\item[Fig. 7] For the $48\times 48$ lattice, along the vertical path with
$r_0 = -2.2$, the results for $m_{sol}$ obtained from
the Kadanoff's operator
and the mass given after the integration of $\Omega(\beta_0)$ are compared.
 We compare them with the classical and
the theoretical perturbative value up to first order in
$O (\lambda_0  / |r_0|)$,
The x-coordinate is  $\beta_0 = {1 \over \lambda_0}$.

\end{itemize}

\end{document}